\documentclass[12pt,preprint]{aastex}

\slugcomment{     }

\shorttitle{Reply to Melott}

\shortauthors{A. B. Romeo et al.}

\begin{document}

\title{Reply to Melott's\\
       Comment on ``Discreteness Effects in Lambda Cold Dark Matter
       Simulations: A Wavelet-Statistical View'' by Romeo et al.}

\author{Alessandro B. Romeo}

\affil{Onsala Space Observatory,
       Chalmers University of Technology,
       SE-43992 Onsala, Sweden}

\email{romeo@chalmers.se}

\and

\author{Oscar Agertz, Ben Moore and Joachim Stadel}

\affil{Institute for Theoretical Physics,
       University of Z\"{u}rich,
       CH-8057 Z\"{u}rich, Switzerland}

\begin{abstract}

   Melott has made pioneering studies of the effects of particle
discreteness in $N$-body simulations, a fundamental point that needs
careful thought and analysis since all such simulations suffer from
numerical noise arising from the use of finite-mass particles.  Melott
(arXiv:0804.0589) claims that the conclusions of our paper
(arXiv:0804.0294) are essentially equivalent to those of his earlier
work.  Melott is wrong: he has jumped onto one of our conclusions and
interpreted that in his own way.  Here we point out the whys and the
wherefores.

\end{abstract}

\section{MELOTT'S CLAIM VS.\ THE FACTS}

\subsection{Our Approach}

   The first point raised by Melott concerns our approach.  He states
that we perform simulations with truncated power spectra to
investigate the effect of initial small-scale power, and that this
approach was introduced by him and collaborators.

   The scope of our paper is much wider than that, and we cite
Splinter et al.\ (1998) among others (cf.\ Sect.\ 1).  We even quote
Melott (2007), a non-refereed astro-ph note submitted as a
comment/review.  And we do that in the central part of the
Introduction, where we motivate our paper and point out what is new:
(1) we assess the actual significance of discreteness effects against
statistical scatter; (2) we offer a deeper and wider view of such
effects through a thorough wavelet-statistical analysis; (3) we probe
particular aspects of the problem, such as the range of scales
affected by discreteness and a variety of statistical effects arising
from the initial conditions.  Thus, when commenting on our approach,
Melott compresses the scope of our paper into a subitem of point (3)
and our 80 simulations into one of the eight subsets!

\subsection{Our Conclusions}

   Melott then comments on our conclusions, which he compresses into a
single sentence: ``They conclude that discreteness effects are visible
in the simulations on all scales $\epsilon<2d$ where $\epsilon$ is the
force resolution (sometimes called `softening'), and $d$ is the mean
interparticle separation''.  He states that our conclusions are
essentially the same as those previously discussed by him and
collaborators.

   Our first conclusion is that dynamical evolution does not propagate
discreteness noise up from the small scales at which it is introduced
(cf.\ Abstract).  This is one important aspect of the robustness of
cosmological $N$-body simulations, which we can prove rigorously
thanks to new and powerful wavelet statistics.  The point is that the
final power spectra, correlation functions and mass variances only
show marginal differences, if any, once their scatter is taken into
account (see Sects 3.1 and 3.2).  This is also true for diagnostics
that are sensitive to the phase of the density fluctuations (see Sects
3.3 and 4.2).  One needs minimum-scatter phase-sensitive statistics,
such as our wavelet set, to show that discreteness noise is not
propagated upwards (see Sects 4.2, 4.3 and 6.1).

   Our second conclusion is that one should aim to satisfy the
condition $\epsilon\sim2d$, where $\epsilon$ is the force resolution
and $d$ is the interparticle distance (cf.\ Abstract).  This condition
involves several aspects of discreteness, which appear if there is
unbalance between force and mass resolution: initial non-Gaussianity
from Gaussian initial conditions, departure from lognormality and rise
of further complexity at low redshifts (cf.\ Sect.\ 7).  This is a
fresh view of the problem, which we discuss together with its
implications.  Concerning the range of scales affected by
discreteness, $\epsilon\la s\la2d$, this is a result that again we can
prove rigorously thanks to our wavelet statistics.  Previous attempts
to quantify this point (e.g., Splinter et al.\ 1998) neglected the
statistical scatter of the diagnostics, which generally dominates over
the systematic effects of discreteness, as pointed out above (cf.\
first conclusion).

\subsection{Our Implications}

   Let us finally remark that the implications of our work are
\emph{NOT} those expressed by Melott, when commenting on cosmological
codes.  \emph{We} conclude that discreteness effects can be kept under
control by implementing our condition $\epsilon\sim2d$ adaptively, not
only in AMR codes but also in tree-based codes, and clarify how (cf.\
Sect.\ 7).

\end{document}